\documentclass[10pt, conference, letterpaper]{IEEEtran}
\IEEEoverridecommandlockouts

\usepackage{cite}
\usepackage{amsmath,amssymb,amsfonts}
\usepackage{algorithm}
\usepackage[noend]{algpseudocode}
\usepackage{graphicx}
\usepackage{xcolor}
\usepackage{tikz}
\usepackage{booktabs}
\usepackage{textcomp}
\usepackage{balance}
\usepackage{url}

\usepackage{makecell}

\usepackage{subcaption}

\def\checkmark{\tikz\fill[scale=0.4](0,.35) -- (.25,0) -- (1,.7) -- (.25,.15) -- cycle;} 

\usepackage{xspace}
\def\scheme{\textsc{F\&F}\xspace}

\usepackage[pdfproducer={}, pdfcreator={}, pdfauthor={}, pdftitle={}, pdfsubject={}, pdfkeywords={}]{hyperref}

\begin{document}

\title{\Large Fast\&Fourier: In-Spectrum Graph Watermarks}

\author{\IEEEauthorblockN{Jade Garcia Bourrée}
\IEEEauthorblockA{\textit{Univ Rennes, Inria, CNRS}\\
Rennes, France
}
\and
\IEEEauthorblockN{Anne-Marie Kermarrec}
\IEEEauthorblockA{\textit{EPFL}\\
Lausanne, Switzerland
}
\and
\IEEEauthorblockN{Erwan Le Merrer}
\IEEEauthorblockA{\textit{Univ Rennes, Inria, CNRS}\\
Rennes, France
}
\and
\IEEEauthorblockN{Othmane Safsafi}
\IEEEauthorblockA{\textit{EPFL}\\
Lausanne, Switzerland
}
}

\date{}

\maketitle

\begin{abstract}
Graphs commonly represent complex interactions, and protecting their provenance helps prevent misuse and preserves trust in the data's source. We address the problem of \textit{watermarking} graph objects, which involves hiding information within them to prove their origin. The two existing methods for watermarking graphs employ subgraph matching or graph isomorphism techniques, which are known to be intractable for large graphs.
To reduce operational complexity, we design a new graph watermarking scheme, \scheme, based on an image watermarking approach that leverages the fact that graphs and images share matrix representations. We analyze and compare \scheme, whose novelty lies in embedding the watermark in the Fourier transform of the adjacency matrix of a graph. Our technique enjoys a significantly lower complexity than that of related works (i.e., in $\mathcal{O}\left(N^2\log N\right)$), while performing at least as well as, if not better than, the two state-of-the-art methods. 
\end{abstract}

\maketitle

\section{Introduction}
\textit{Watermarking} is the art of hiding information within a digital object. Watermarking has proven to be efficient in embedding robust information in complex data~\cite{hartung1999multimedia, kumar2020recent}. Despite the explosion in the use of graph-structured data in various scientific fields~\cite{bonifati2022special,das2014tale,de2014r2g,cudre2011graph}, the research of \textit{graph watermarking} --- especially for unweighted graphs --- is still in its early stages~\cite{isc, COSN}.

Creating graphs requires extensive data collection and pre-processing, making them highly valuable assets. Graphs emerge from a wide range of applications, including webs of trust in blockchains, circles in social networks, and co-purchasing networks in e-commerce platforms. The creators of these graphs often seek recognition, similar to the practices encouraged by Creative Commons licences. Watermarking schemes meet this demand by embedding provenance information as a digital signature, which the graph owner can extract later to prove ownership.

Despite its potential, the watermarking of graph objects remains uncommon. None of the popular online platforms that share graph representations~\cite{snapnets, konect, nr} currently propose watermarked graphs. The main reason is the limited number of existing techniques, which are too computationally expensive. These techniques rely on intensive graph-related operations, such as subgraph matching and isomorphism~\cite{isc, COSN}.

In this paper, we introduce a competitive watermarking scheme inspired by image watermarking techniques~\cite{TIP}. By treating adjacency matrices of graphs analogously to pixel matrices in images, we propose to leverage complexity-reduced computational operations for watermarking graph adjacency matrices. The appendix~\ref{sec:related} presents an overview of methods for watermarking graph-related objects. Our contributions are:

1) We introduce a novel analytical framework that applies an image processing scheme directly to the adjacency matrices of graphs, to use operations faster than the NP-complete related complexities of the related works \cite{isc, COSN}.

2) The main technical challenge in applying an image watermarking technique in a graph context is related to the binary nature of an edge's presence in an adjacency matrix, as opposed to real numbers in an image matrix. This forces binarization. Our formal analysis examines the necessary binarization of the watermark to verify that it effectively results in a unique watermark in the graph (i.e., without collision with another watermark insertion), thereby setting the stage for subsequent practical applications.

3) Finally, we perform a head-to-head comparison of \scheme with the related work in the graph domain (schemes by Zhao \textit{et al.}~\cite{COSN} and Eppstein \textit{et al.}~\cite{isc}). We conclude that \scheme performs at least as well as its competitors, with the advantage of being much faster.

\section{Goals and Threat Model}
\label{sec:goals}
 
\subsection{Watermarking Graphs: Threat Model}

A graph watermarking scheme is a set of functions (\texttt{Keygen}, \texttt{Embed}, \texttt{Extract}), applied to a graph~\cite{isc}. With \texttt{Keygen}, the owner of a graph generates a secret key. This key is embedded into her graph $G$ with the \texttt{Embed} operation. It produces a modified graph $G_W$ and a watermark $W$ (which is related to the difference between the original and the watermarked graph). The graph's owner can prove her ownership of a suspected shared graph $G^*$ with the watermark $W$ using the \texttt{Extract} function. \texttt{Extract} returns a success if $G^*$ is watermarked with $W$. Otherwise, either 1) $G^*$ does not contain the watermark $W$ embedded using the considered scheme (or does not contain any watermark), or 2) an attacker has modified $G^*$ sufficiently to prevent \texttt{Extract} from succeeding.

\subsection{Design Goals for Watermarking Schemes}
\label{ss:goals}

The goals of a watermarking scheme are as follows~\cite{COSN}: 
\paragraph*{(Goal 1) Low distortion}
The watermark embedding with \texttt{Embed} must have a low impact on the original graph to preserve the intrinsic value of that graph.

\paragraph*{(Goal 2) Watermark uniqueness}
The \texttt{Embed} function must be injective: a graph watermarked independently with two different keys should not produce two identical graphs.

\paragraph*{(Goal 3) False positives and negatives}
A good watermarking scheme must minimize false positives without generating too many false negatives. 

\paragraph*{(Goal 4) Robustness to modifications}
A watermarking scheme must be resistant to modifications (i.e., attacks) performed on the watermarked graph. More specifically, the key extraction must operate with a high probability, even in the presence of attacks.\\

Other goals, such as undetectability and distortion effect, are defined and studied in Appendix \ref{a:add-exp}. We have established a framework based on the state-of-the-art watermarking methods.

\section{\scheme: Cox \textit{et al.} Scheme for Graphs}
\label{sec:algo}

This work considers undirected and unweighted graphs, to compete with state-of-the-art schemes~\cite{isc, COSN} described in Section~\ref{ss:Zhao-Eppstein}. Since the goal is to exploit watermarking schemes designed initially for images, we assume that we can label the input graph arbitrarily, so that a deterministic mapping of its vertices in the adjacency matrix can be performed~\cite{fekete2015reorder,hahsler2011dissimilarity}. This assumption is realistic in many cases where vertices are naturally labeled, and it is consistent with the fact that for images, pixels have a deterministic position in their matrix representation.

\subsection{Framework Rationale}
Table~\ref{notations} and Figure~\ref{fig:schema} sum up the notations introduced hereafter (Appendix~\ref{s:notations}). The number of vertices of a graph $G=(V, E)$ is $N = |V|$. The adjacency matrix $A$ of an unweighted graph $G$ is a binary square matrix of dimension $N$x$N$: there is a $1$ in $A[i,j]$ if the edge $(i,j)$ exists, otherwise $0$. The process treats this matrix as an image. Any image-based watermarking scheme that uses a Fourier transform can watermark $A$~\cite{TIP,joseph1998ruanaidh,pereira1999template,8918869,riaz2008invisible,soni2013image}. Thus, we use the Cox \textit{et al.} scheme~\cite{TIP} as it is a pioneering work. From this scheme, we obtain a matrix $A'$ that contains a watermark. $A'$ has the salient property of being composed of real numbers (rather than of binary values, such as in $A$). $A'$ is also no longer symmetric. 

Therefore, our framework includes additional steps to transform $A'$ back into an undirected and unweighted graph. A key part of our challenge is to analyze the consequences of these steps.
More specifically, the required binarization involves thresholding all values in $A'$ by the average of $A$. Let $av(A)$ denote this real value. Each element becomes $0$ if its modulus is less than the average of the original adjacency matrix $A$; otherwise, it becomes $1$.

Our framework yields a proper graph representation through its new adjacency matrix $A_W$. From this matrix, the resulting graph can be shared at will by the owner of the graph. \\

\par Next, we describe the original scheme of Cox \textit{et al.} for watermarking images.

\subsection{Watermarking Images with the Cox \textit{et al.} Scheme}
In~\cite{TIP}, the watermark is a sequence of a fixed number of random reals chosen from a Gaussian distribution. The authors compute the Fourier transform of the original image and insert the watermark into the most significant low-frequency coefficients of the transformed matrix. They propose three ways to insert the watermark into the highest magnitude coefficients, including a simple additive method.

Starting from a possibly altered image, the procedure extracts a watermark in the spectral domain, compares it to the original, and computes a similarity score to decide if the image contains the watermark.

\subsection{The \scheme Scheme for Watermarking Graphs}

We now propose \scheme as an adaptation of the Cox \textit{et al.} scheme for unweighted graphs with pseudo-code Algorithm~\ref{algo}. We discuss the main differences with the original Cox \textit{et al.} scheme.

\begin{algorithm}
\caption{\scheme scheme}
\begin{algorithmic}[1]
 {\small
 
 \Function{Keygen}{$m, \sigma$}
 \State $\omega = \left [ \mathcal{G}(0,\sigma^2), \ldots , \mathcal{G}(0,\sigma^2) \right ]$ \Comment{
 of length $m$}
 \State \textbf{Return} $\omega$
 \EndFunction}

 {\small
 \Function{Embed}{$ A, \omega$}
 \State $m = |\omega|$
 \State $th = av(A)$ \Comment{
 real number}
 \State FT(A) = Fourier transform of $A$
 \State $\chi = \texttt{argsort}(|FT(A)|)[0:m]$ \Comment{ 
 \texttt{argsort} returns the indices that would sort its argument}
 \State $\mathcal{W} [i, j] = \left . \begin{matrix}
 \omega [n] \text{ if there exists } n \leq m, \; \chi[n] = (i,j)\\
 \end{matrix}\right.$
 \State $A' = A + FT^{-1}(\mathcal{W})$
 \State $A_W [i, j] = \left \{ \begin{matrix}
 1 \text{ if } |A'[i,j]|>th\\ 
 \text{else } 0
 \end{matrix}\right.$ 
 \Comment{
 binarization}
 \State \textbf{Return} $A_W$
 \EndFunction
 }

 {\small
 \Function{Extract}{$A$, $A^*, \omega, \theta$}
 \State $A_W = ...$ \Comment{
 generated as in \texttt{Embed}}
 \State $W^* = FT(A - A^*)$
 \State $W = FT(A - A_W)$
 \State $s = \left\|W^* - W \right\|_2$
 \State \textbf{Return} $s \leq \theta*\left\| W \right\|_2$ \Comment{
 True: watermark retrieved}
 \EndFunction 
}
\end{algorithmic}
\label{algo}
\end{algorithm}

\paragraph*{Key generation}
\texttt{Keygen}$(m, \sigma)$ generates a key as a Gaussian vector of $m$ values with standard deviation $\sigma$. This step is identical to that of Cox \textit{et al.}.

\paragraph*{Watermark embedding} The graph owner wants to embed a key $\omega$ generated with the function \texttt{Keygen} into her graph $G$ thanks to its adjacency matrix $A$. First, the sequence $\chi = \{(i_1, j_1), \dots, (i_m, j_m)\}$ is computed as the sequence of the first $m$ indices that would sort the modulus of the Fourier coefficients of $A$. The process relies on the sequence $\chi$ to compute the intermediate watermark $\mathcal{W}$ of the graph owner in the Fourier spectrum. $\mathcal{W}$  is the matrix representation of the key where each element $\omega[n]$ is at position $\chi[n] = (i_n, j_n)$. Then, the Fourier inverse of the intermediate watermark  $\mathcal{W}$  is added to the adjacency matrix $A$ to obtain a real matrix $A'$. The algorithm continues with binarization and symmetrization to obtain $A_W$. The watermark $W$ is equal to the difference of all coefficients of the Fourier transform of $A$ and $A_W$:
$ W = FT(A) - FT(A_W).$

\paragraph*{Watermark extraction}
The suspected watermark $W^*$ is compared to the original watermark $W$ using the $2$-norm of their difference. If the difference between $W^*$ and $W$ is less than a threshold, the extraction is hypothetically successful.

\paragraph*{The specificity of \scheme}
The first difference between the original scheme of Cox et al.~\cite{TIP} and \scheme concerns the watermark key, which depends on the binarization operation rather than allowing arbitrary selection. What is left after binarization depends on the chosen length $m$ and the standard deviation $\sigma$ of the elements in that key. Recommendations in Cox \textit{et al.}~\cite{TIP} are no longer valid due to this binarization step. Section~\ref{ss:distorsion} discusses how to obtain a key properly.

The second specificity is also due to the binarization step. This step spreads the key across the entire transformed matrix. As a consequence: $\mathcal{W} \neq W.$
Thus, the watermark is defined over all matrix coefficients and not only over the ones specified by $\chi$. The function \texttt{Extract} is applied to the watermark $W$ resulting from the binarization after embedding $\omega$ with the Cox \textit{et al.} scheme, and \textbf{not} applied directly to the key $\omega$ in the coefficients specified by $\chi$.
The procedure uses the key $\omega$ to derive $W$, either by computing it through the embedding function or by calling \texttt{Extract}, then comparing it to $W^*$.

\subsection{Scheme complexity vs related works}\label{ss:complexity} The main advantage in applying an image watermarking technique in a graph context is the time complexity of the watermarking scheme, which uses vertex labels to have a consistent mapping in the adjacency matrix. The function \texttt{Keygen} runs in constant time in $m$ operations, where $m$ is the length of the desired key, as in Cox \textit{et al.} In \texttt{Embed}, the worst-case time complexity of the Fourier operations (l.7 and l.10) and the sorting operation to compute $\chi$ (l.8) are linearithmic in the number of coefficients of their inputs. In other words, they require $\mathcal{O}\left(N^2\log N\right)$ operations, as each applies to an $N$x$N$ matrix. The time complexity of all other operations in \texttt{Embed} are at most linear in the number of coefficients of their inputs. That is to say, the worst-case time complexity of \texttt{Embed} is $\mathcal{O}\left(N^2\log N\right)$. The function \texttt{Extract} uses the same operations as in the function \texttt{Embed}, with the addition of computing the $2$ norm of the difference between two matrices of dimension $N$x$N$, which is linear in the size of the object. The worst-case time complexity of \texttt{Extract} is thus also $\mathcal{O}\left(N^2\log N\right)$.

For comparison, Zhao \textit{et al.} use an NP-complete subgraph matching routine and suggest lower complexity on graphs with "very high node degree heterogeneity", though without formal analysis (see \cite{COSN}, Section 4); the problem remains NP-complete in general. Eppstein \textit{et al.} \cite{isc} prove correctness for Erdős-Rényi and power-law graphs but provide no bounded complexity analysis, so their scheme also remains NP-complete. Section~\ref{sss:timings} compares all three schemes by execution time.

\section{Experiments: Compliance to Design Goals}
\label{sec:experiments}

This section presents the results of an extensive experimental study of \scheme.
The structure highlights how \scheme meets the design goals outlined in Section~\ref{ss:goals}.

\subsection{Experimental Setup}
Using Python 3, we conduct experiments and record timings from runs on a server featuring Intel Xeon E5-2630 v3 CPUs operating at 2.40 GHz.

\paragraph*{Graph generative models}
For experiments involving variations in graph properties, we use graph generators from the Python package NetworkX~\cite{hagberg2008exploring}. These generators implement well-known models applied to various real-world graphs, such as the Erdős-Rényi (ER), Barabási-Albert (BA), and Watts-Strogatz (WS) models.

\paragraph*{Large and real graphs}
We also conduct experiments with \scheme on eight real-world large graphs with varying edge densities and different structures, obtained from the SNAP repository~\cite{snapnets} (e.g., Pokec) and the Network Repository~\cite {nr} (e.g., Flickr).

\paragraph*{Evaluation metrics} The properties and information to preserve are highly context-dependent, depending on the intended use of a graph. It is impractical to cover all possible scenarios. Due to its generality and widespread use in the state-of-the-art~\cite{isc, COSN}, we consider the graph \textit{edit distance} to quantify the degradation of a graph:

\paragraph*{Definition} The graph \textit{edit distance} (denoted ED) is the percentage of edges that distinguish two graphs. In the sequel,  the distance refers to the comparison with the original graph.

For instance, if the $embed$ function flips $200$ edges from a graph $G$ containing $2,000$ edges, then the distance based on the fraction of edges between $G'$ and $G$ is $10$ \% edges. 
We emphasize that the ED can be greater than $ 100$ \% because we can add edges, not just remove them. One can flip at most $N(N-1)/2$ edges from the input graph.

\subsection{Adaptations from the Theoretical Scheme}\label{ss:exp}
\subsubsection{\scheme Scalability}
\label{ss:scalability}

To increase the scalability of their schemes when applied to large graphs, both related works~\cite{isc, COSN} filter out the input graph to operate on a smaller set of vertices. Inspired by the Cox \textit{et al.} scheme~\cite{TIP} \scheme also performs a dimensionality reduction. This step is described in detail in the Appendix~\ref{a:reduction}.

\subsubsection{Resilience to Attacks and the Similarity Threshold}\label{sss:theta}
As in the original Cox \textit{et al.} scheme and in the related work~\cite{COSN} and~\cite{isc} for graphs, there is a \textit{threshold} parameter $\theta$ driving the success of the extraction function in Algorithm~\ref{algo}. The following paragraphs deal with this parameter $\theta$.
\paragraph*{The ideal case: no attack on the watermarked graph} We show that when attacks on $G_W$ do not occur, setting $\theta = 0$ guarantees the successful extraction of the graph owner's watermark. In other words, no false negatives arise for unattacked watermarked graphs.

\par \textit{Proof.} A false negative occurs when \texttt{Extract} fails despite the tested graph containing the given watermark. Specifically, the extraction fails only if the following condition does not hold:
 $\left\|W^* - W \right\|_2 \leq \theta*\left\| W \right\|_2,$
where $W^*$ and $W$ are computed according to Algorithm~\ref{algo}, and while the tested graph $G^*$ \textbf{is} the watermarked graph $G_W$, that has not been attacked. Thanks to the deterministic mapping of vertices to adjacency matrices, $A^* = A_W$. Since the Fourier transform is linear, the condition $\left\|W^* - W \right\|_2 \leq \theta*\left\| W \right\|_2,$ becomes: $0 \leq \theta*\left\| W \right\|_2$, which always holds since $\theta \geq 0$. Thus, if $\theta = 0$, no false negatives can occur with the \scheme scheme.

\paragraph*{A realistic case with an attacked watermarked graph} In practical scenarios, the owner shares a watermarked graph that may undergo modifications. A false negative occurs when \texttt{Extract} fails despite the graph containing the owner's watermark. Here, $\theta > 0$ sets the similarity threshold above which the owner accepts the extracted watermark as sufficiently close to the original, despite alterations.

We set $\theta$ to tolerate $10$\% edge flips, which is a conservatively high value (according to related work~\cite{isc, COSN}, as it will be illustrated in Section \ref{ss:distorsion}). Consequently, no false negatives can occur with less than $10$ \% edge flips. We study the effect of $\theta$ on false positives in Section~\ref{ss:FPFN}.

\subsection{(Goal 1) Low Distortion}\label{ss:distorsion}
According to related work~\cite{isc, COSN}, the ED of large graphs must be at most $10^{-2}$\% edges. We now present a method for automatically setting parameters suitable for watermarking with minimal distortion. We study it theoretically in Appendix~\ref{s:theory}.

Figure~\ref{fig:existence} illustrates the link between $\sigma$ and $m$ for our three generative graph models, and a slight distortion of $0.005$\% edges.

Key lengths used in~\cite{isc} depend on \textit{densities} (being defined as the number of edges over the number of vertices) \cite{melancon2006just}, while those of~\cite{COSN} depend on the number of vertices. We set the key length $m$ of \scheme as the key length used in~\cite{isc} because the experiment in Section~\ref{exp:unique} shows that \scheme is also related to densities.

Given the key length $m$ defined as in~\cite{isc}, it is always possible to find a value of $\sigma$ to watermark the graph while achieving the distortion goal. Experiments have shown that for any graph $G$ and any fixed key length $m \in [\![1, N_0*(N_0-1)/2]\!] $, there exists a value $\sigma_{max}$ to watermark the graph with a strictly positive edit distance. An automated way to set $\sigma$ under the small ED constraint is to use a dichotomous search in $[\![1, \sigma_{max}]\!]$. As shown in Appendix~\ref{a:reduction}, the proposed guidelines do not affect the scalability of \scheme.

\begin{figure}[t!]
\centerline{\includegraphics[width=0.7\linewidth]{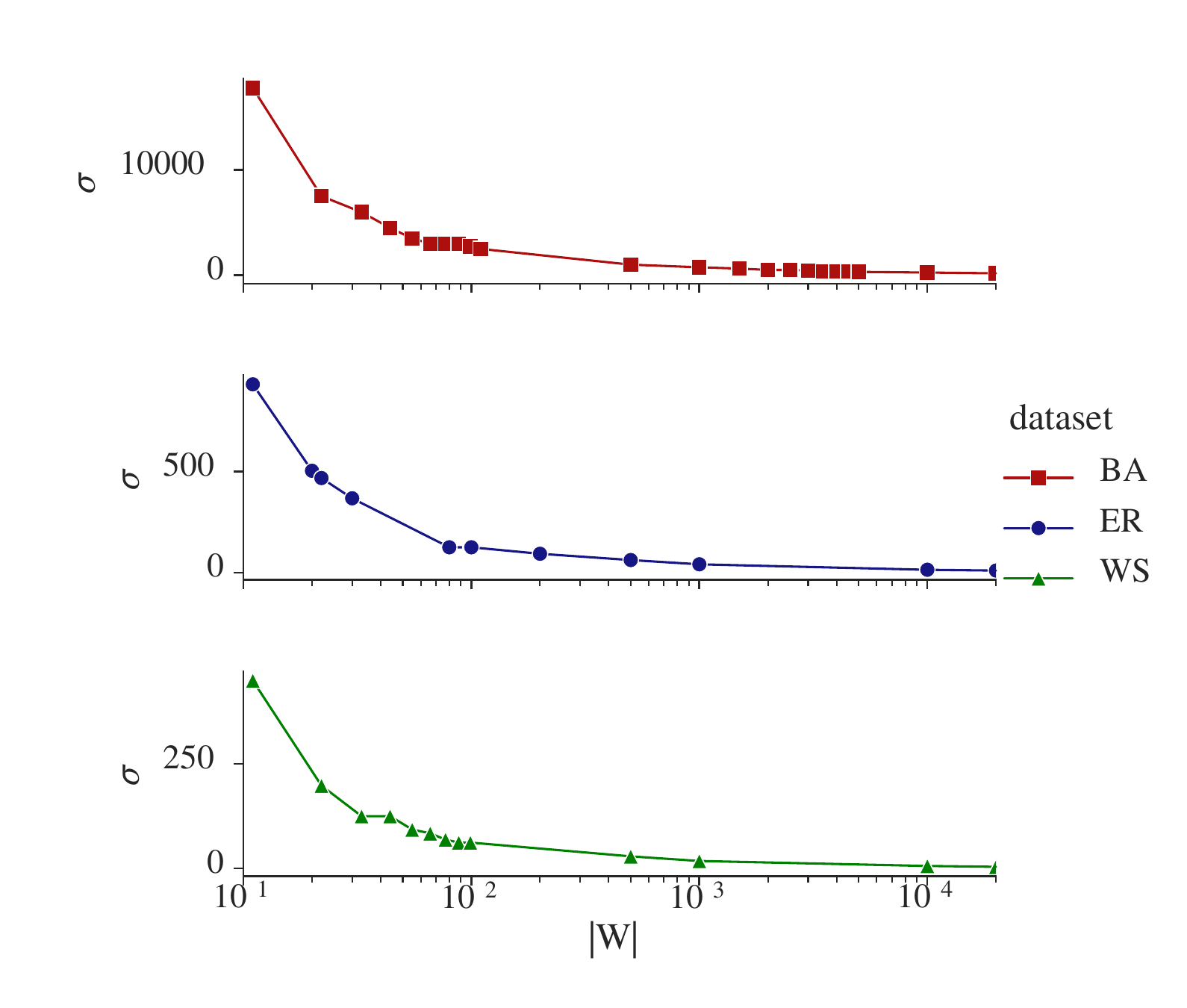}}
\caption{The relation between $m$ and $\sigma$ to watermark graphs with an ED $\in ]0,0.005[$ \% of edges, for the three graph models.}
\label{fig:existence}
\end{figure}

\begin{table}[ht!]
 \begin{center}
 \begin{tabular}{|c|c|c|c| }
 \hline
 Graph & $m$ & $\sigma$ & Watermarking ED\\
 \hline
 BA & $210$ & $1,750$ & $3.10^{-8}$\\
 Flickr & $3,250$ & $32,000$ & $2.10^{-8}$\\
 Pokec & $170$ & $7,000$ & $5.10^{-9} $\\
 \hline
 \end{tabular}
 \end{center}
 \caption{Parameters and resulting distortion (ED).}
 \label{keysetting}
 \end{table}

Applying this method to the three significant real graphs results in the parameters listed in Table~\ref{keysetting}. The key length mmm varies significantly according to each graph's inner topological characteristics. The observed edit distances remain well under the $10^{-2}$\% target, fulfilling Goal 1.

Further experiments on the impact of \scheme on graphs appear in Appendix~\ref{a:add-exp}.

\subsection{(Goal 2) Uniqueness of Watermarks}
\label{exp:unique}

The uniqueness property resulting from \scheme is assessed by generating two keys $\omega_1$ and $\omega_2$ with the same parameters ($m$ and $\sigma$) found by the automatic dichotomous procedure. There is a lack of uniqueness if the two keys result in the same watermark (i.e., if $\omega_1 \neq \omega_2$ and $G_{W_1}=G_{W_2}$).

\begin{figure}[ht]
\centerline{\includegraphics[width=0.7\linewidth]{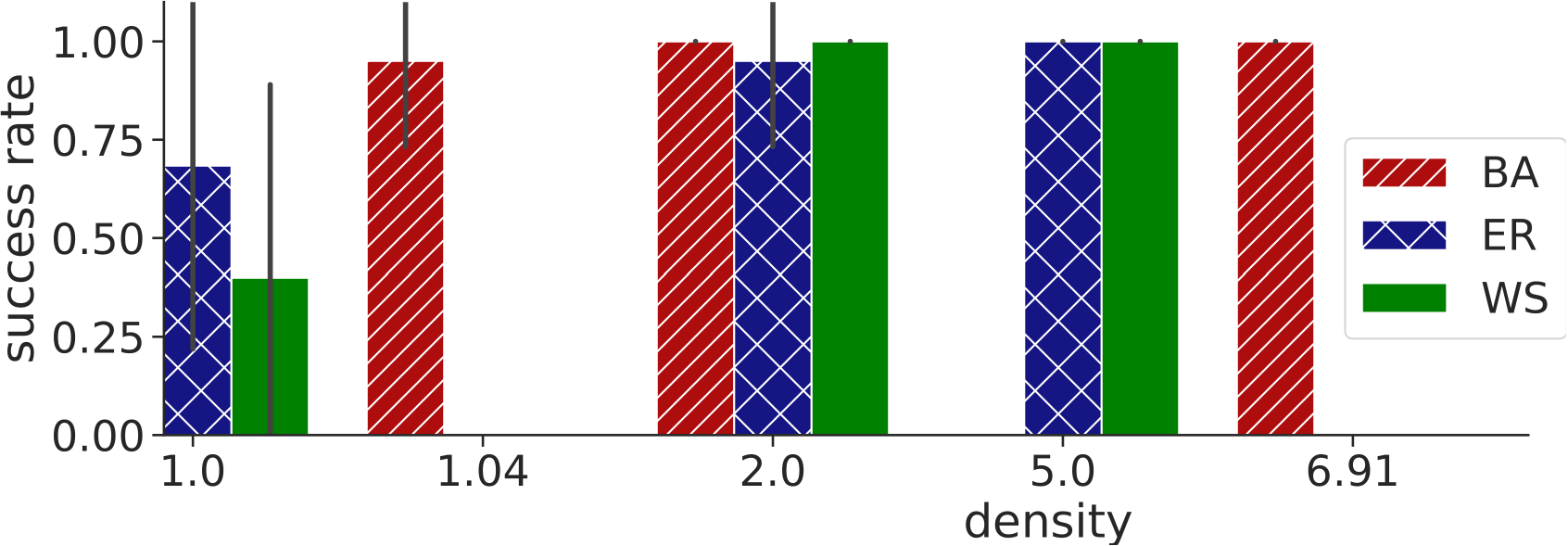}}
\caption{Success rate ($y$-axis) in inserting a unique watermark in three graph models, for 1M vertices graph models and different densities on the $x$-axis.
}
\label{fig:uniqueness}
\end{figure}

We experiment with uniqueness for the three graph models and various graph densities in Figure~\ref{fig:uniqueness}. There are some collisions for the lowest densities, beyond $2$. 
It is not an issue per se, as these graphs are rare or atypical: a density less than $2$ means that a graph has, or is close to having, disconnected components. Above this threshold, there is no collision; this is the expected setup, since collected graphs generally have high densities (\textit{e.g.,} from 12 to 36 in web crawls \cite{melancon2006just}, or without density limit for graphs studied in node classification, where edges are even removed to reduce the density before processing~\cite{li2022graph}). It thus fulfills Goal 2 for practical setups.

\subsection{(Goal 3) Low False Positives and Negatives}\label{ss:FPFN}
We now study the effect of $\theta$ on false positives. Let $G_W$ be the watermarked graph of $G$ with the key $\omega$, and $G^*$ be any other graph \textbf{not} watermarked with $\omega$. Recall that Algorithm \ref{algo} with $G^*$ results in a false positive if the extraction of $W$ in $G^*$ succeeds.

\begin{figure}[ht]
\centerline{\includegraphics[width=0.6\linewidth]{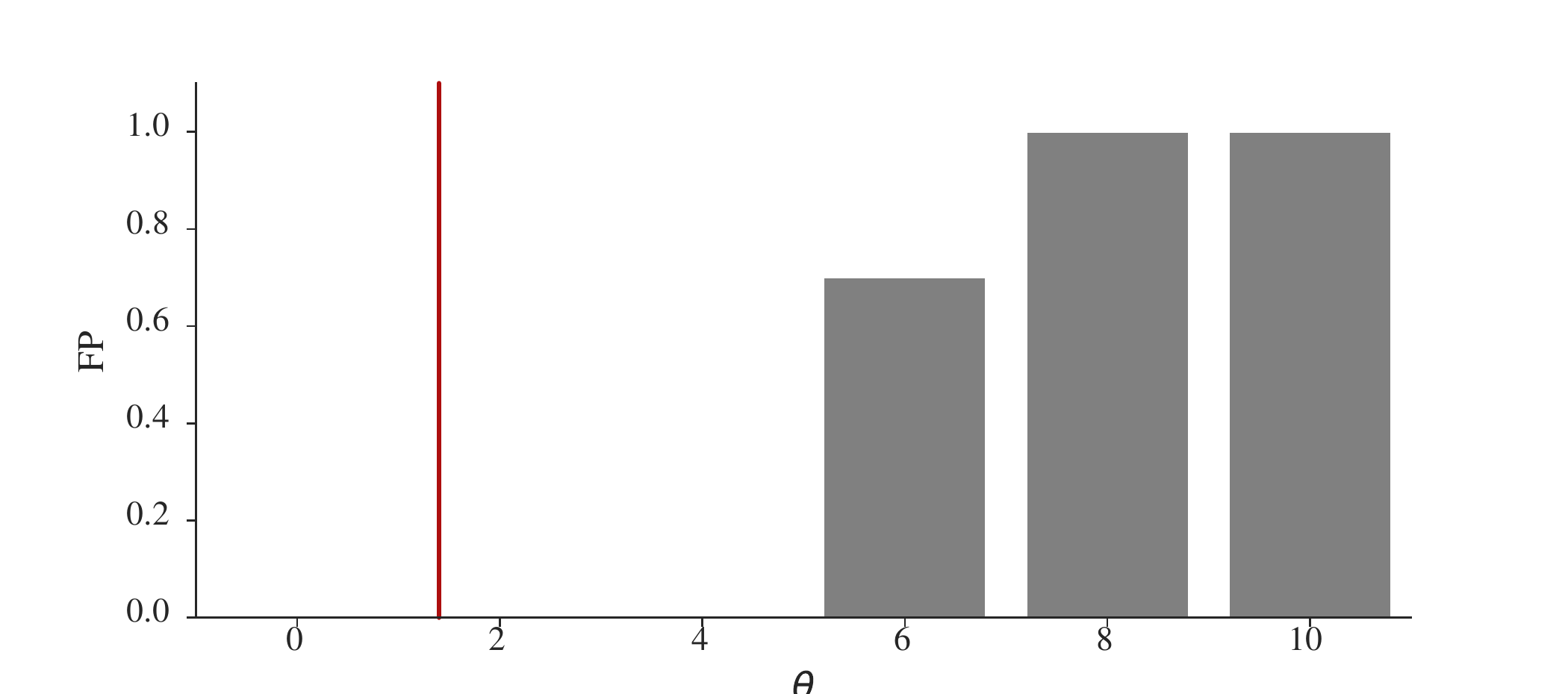}}
\caption{The impact of $\theta$ on extraction false positives (BA graphs).}
\label{fig:choiceT}
\end{figure}

The goal of the following experiment is to determine the value of $\theta$ for which false positives occur when attempting an extraction on non-watermarked graphs. We generate ten BA graphs with density $\frac{|E|}{|V|} = 5$ (according to the uniqueness experiment in Figure~\ref{fig:uniqueness}). For each of them, we first inserted a key $\omega$; then we regenerated five more graphs with the same generative parameters, and tried to extract $\omega$ from them. Extractions occur for different values of $\theta$. Figure~\ref{fig:choiceT} shows that as long as $\theta$ is less than or equal to $5$, there are no false positives. Then, values of $\theta$ above $6$ introduce some. Thus, $\theta = 1.4$ is consistent with the $2$-norm of the watermark to tolerate $10$ \% edge flips for a graph of density $5$ in BA models. Goal 3 is then achievable by parameterization.

In the remaining experiments, $\theta$ allows up to $10$ \% edge flips since no false positives or false negatives occur in this setup.

\subsection{(Goal 4) Robustness to Attacks}

\subsubsection{Intensity and Impact of the Attack}
\label{ss:impact}
\begin{figure}[h!]
\centerline{\includegraphics[width=.8\linewidth]{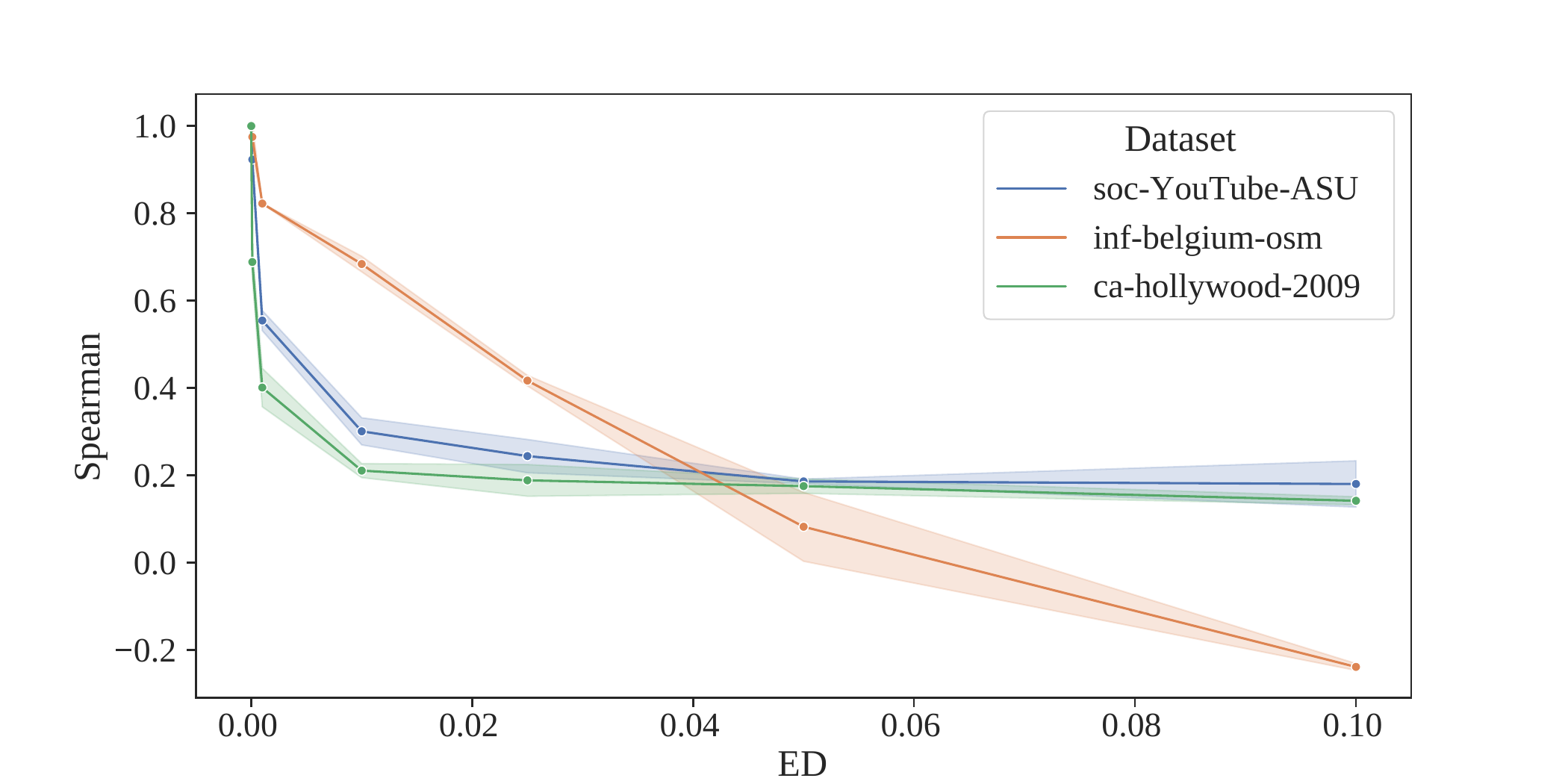}}
\caption{The effect of an edge flipping attack on three real graphs, regarding the degree centrality of the top-1000 most central nodes. The \% of edges flips appears on the $x$-axis, while the Spearman correlation of the top-1000 ranking appears on the $y$-axis.}
\label{fig:spearman}
\end{figure}

To assess a plausible intensity of an attack that a graph can sustain without losing its utility, we show in Figure~\ref{fig:spearman} a simple function performed on the three first graphs of Table \ref{expres}: the ranking of the top-$1000$ nodes according to the \textit{degree centrality}~\cite{das2018study}. The figure plots the changes in this ranking of the most central nodes, according to the classic Spearman metric~\cite{diaconis1977spearman}. It measures the correlation between two rankings of the top nodes (on the original graph and the attacked version). 
The correlation drops dramatically as the attack increases, even with a low value of $1$ \% of edges flipped. Thus, this random attack disrupts the connectivity of the central nodes, casting doubt on the usefulness of the resulting graph. In this context, $10$\% of flipped edges is an extreme attack that appears to remove the utility of the graph, and can thus be considered a very conservative upper bound.

\subsubsection{Robustness}
Table~\ref{expres} reports the performance of \scheme on large real graphs. The key lengths align with those used by Eppstein et al.~\cite{isc}, while standard deviations produce watermarks with minimal distortion. Despite varying densities, all graphs received watermarks that caused only minor distortions, with edit distances ranging from $10^{-6}$ to $10^{-8}$, and reaching as low as $10^{-10}$ for the "inf-belgium-osm" graph.

\begin{table}[ht]
\begin{center}
\begin{tabular}{|c|c|c|c|c|c| }
 \hline
 \makecell{Graphs\\ ($|V|$, $|E|/|V|$)} & $m$ & $\sigma$ & ED & $\theta$ & \scheme\\
 \hline
 \makecell{inf-belgium-osm \\ (1.4M, 1.1)} & 54 & 250 & $3.10^{-10}$ & 0.6 & \checkmark\\
 \makecell{soc-YouTube-ASU \\ (1.1M, 2.6)} & 200 & 12k & $1.1.10^{-8}$ & 2 & \checkmark\\
 \makecell{hollywood-2009 \\ (1.1M, 52.7)} & 162 & 224k & $1.6.10^{-8}$ & 205 & \checkmark\\
 \makecell{rgg-n-2-20-s0 \\ (1M, 6.6)} & 119 & 1,624 & $3.8.10^{-8}$ & 10.2 & \checkmark\\
 \makecell{kron-g500-logn20 \\ (1M, 42.6)} & 71128 & 16k & $3.5.10^{-8}$ & 374 & \checkmark\\
 \makecell{scale21-ef16-adj \\ (1.2M, 51)} & 44,743 & 12k & $1.7.10^{-8}$ & 233 & \checkmark\\
 \makecell{roadNet-PA \\ (1.1M, 1.4)} & 48 & 184 & $6.10^{-6}$ & 0.6 & \checkmark\\
 \makecell{delaunay-n20 \\ (1M, 3.0)} & 63 & 61 & $4.3.10^{-8}$ & 6.2 & \checkmark\\
 \hline
\end{tabular}
\end{center}
\caption{Experiment settings and results for \scheme, on real large graphs.}
\label{expres}
\end{table}

We now subject these graphs to attacks and observe the resilience of \scheme. The last column shows that the extractions are successful despite the attack. We also note that additional tests have shown that all watermarks can also be extracted even under an attack that flips $100$ \% edges, for a value of $\theta$ set to be resistant to only $10$ \% edge flips. It underscores the robustness of \scheme, and the reach of Goal 4.

This experimental section confirms both the applicability of \scheme to real graphs and the low distortion of the scheme on these graph structures, as well as its resilience to significant attack strengths.

\subsubsection{Densities and the Resulting Threshold}
\begin{figure}[h!]
\centerline{\includegraphics[width=\linewidth]{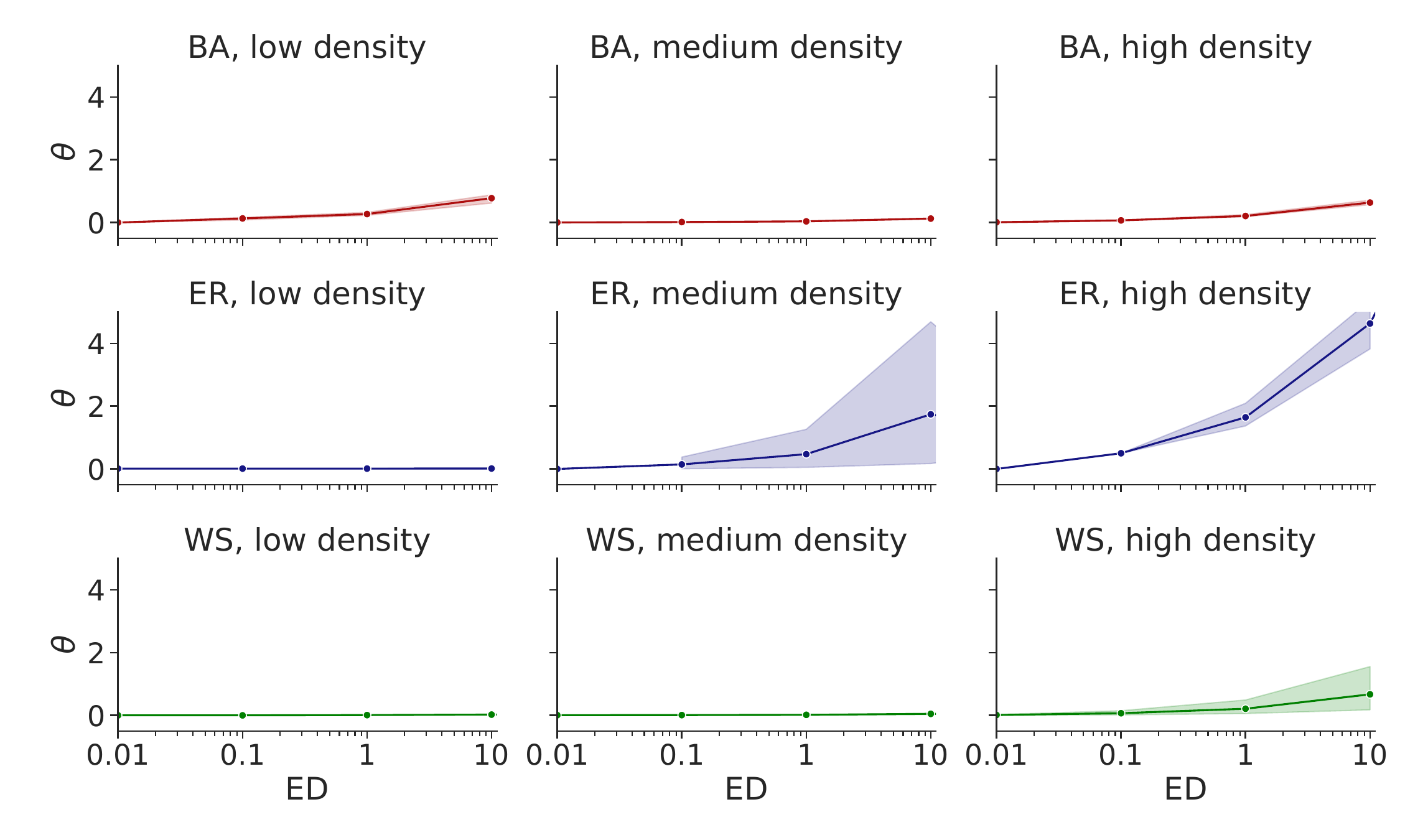}}
\caption{The choice of $\theta$ in \scheme to be resistant to $x$\% edge flips depending on the density and the type of  graph model.}
\label{fig:Tdensitymodel}
\end{figure}

The final experiment for \scheme is to test its robustness to different graph densities across the three graph models. We take the three graph generators (BA, WS, ER) and vary the density $\frac{|E|}{N}$ of graphs to $5$, $30$, and $50$. $5$ is a low density which respects the uniqueness condition, $30$ and $50$ are medium and high densities according to~\cite{melancon2006just} and the graphs taken from SNAP~\cite{snapnets}. For each generator and each density, we set $\theta$ to be resistant to $x$ \% edge flips ($x$-axis). This resilience is observed in $3$ runs per point on Figure~\ref{fig:Tdensitymodel}, where a dichotomous search sets $\theta$ to achieve the effective resilience to the attack.

Figure~\ref{fig:Tdensitymodel} shows that the resulting threshold $\theta$ depends primarily on the type of graph. For BA or WS graphs, a setting $\theta = 1$ is sufficient for \scheme to be resistant to attacks with up to $10$\% edge flips ($x$-axis), regardless of the density of the original graph. Density is the most relevant parameter when watermarking ER graphs, where $\theta$ can vary from $0.3$ to $5$ to be resistant to attacks of up to 10\% edge flips.

\section{A Comparative Benchmark}
\label{sec:benchmark}
Finally, we perform a head-to-head comparative benchmark of \scheme with the two state-of-the-art approaches~\cite{COSN} (Zhao) and~\cite{isc} (Eppstein), both in terms of running time and resilience to attacks.

\subsection{Related work}\label{ss:Zhao-Eppstein}

\paragraph*{The scheme by Zhao \textit{et al.}~\cite{COSN}} There are four steps in the graph embedding function of Zhao \textit{et al.}~\cite{COSN}. First, they generate a random seed based on cryptographic keys. Then, they generate a random watermark graph (Erdős-Rényi) using the seed and match it with a subgraph of $G$. The selection of the subgraph relies on the structure of the vertices (i.e., their neighborhoods). Finally, an XOR operation performs the embedding between the subgraph and the random watermark graph. For the extraction, they regenerate the watermark graph and the subgraph, just as in the embedding process. After that, they identify in the potential $G^*$ which vertices are candidates to match the subgraph (again using the vertex structure). Finally, they extract the watermark using a recursive algorithm. Since the authors did not provide the code, we implemented it in Python 3. We note that our execution results are consistent with those presented by the authors in their experimental section (Figure 2b in \cite{COSN}). We set $\theta=0.95$ and $B=25$, for an acceptable execution time.

\paragraph*{The scheme by Eppstein \textit{et al.} ~\cite{isc}} The scheme of Eppstein \textit{et al.}\cite{isc} consists of the same three functions \texttt{Keygen,} \texttt{Embed} and \texttt{Extract} as the \scheme scheme. In the scheme of Eppstein \textit{et al.}, private keys generated by the \texttt{Keygen} function are sets of vertex pairs in a graph containing as many vertices as in the graph to watermark. There are three main steps in the embedding function. The first step is to order and label the vertices of the graph, taking into account their degrees and using bit vectors. The vertices fall into three groups: high-degree, medium-degree, and low-degree vertices. Then, the procedure randomly flips each edge included in the private key according to the probability of its statistical existence in the graph. The function \texttt{Extract} uses graph isomorphism to identify the watermarked graph. Unlike \scheme and Zhao~\cite{COSN}, this method identifies a graph between multiple (simultaneously created) watermark copies, not just a single watermark graph with a given key. The selected copy is the closest copy considering the Hamming distance.

\subsection{Execution Timings}
\label{sss:timings}

We begin by benchmarking the runtime of the \texttt{Embed} and \texttt{Extract} functions from the competing watermarking schemes. Keys follow the generation guidelines from each paper, with timing results excluded because they execute very quickly. Implementations include scalability enhancements suggested by the original works.

\begin{figure}[t!]
 \begin{subfigure}[t]{0.5\textwidth}
 \centering
 \includegraphics[height=2.1in]{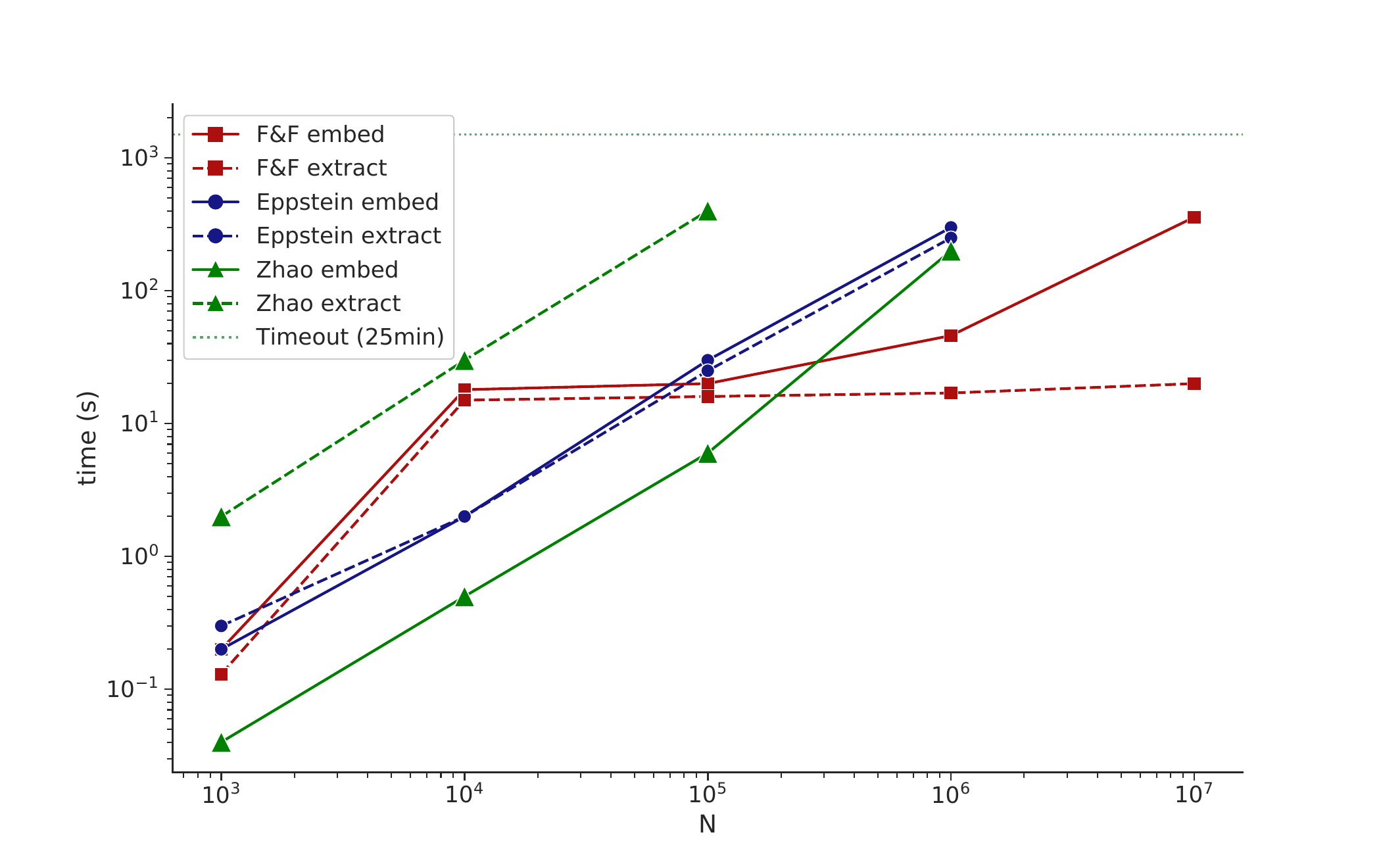}
 \caption{Execution timings\label{fig:timings}}
 \end{subfigure}%
 
\begin{subfigure}[t]{0.5\textwidth}
 \centering
 \includegraphics[height=2.7in]{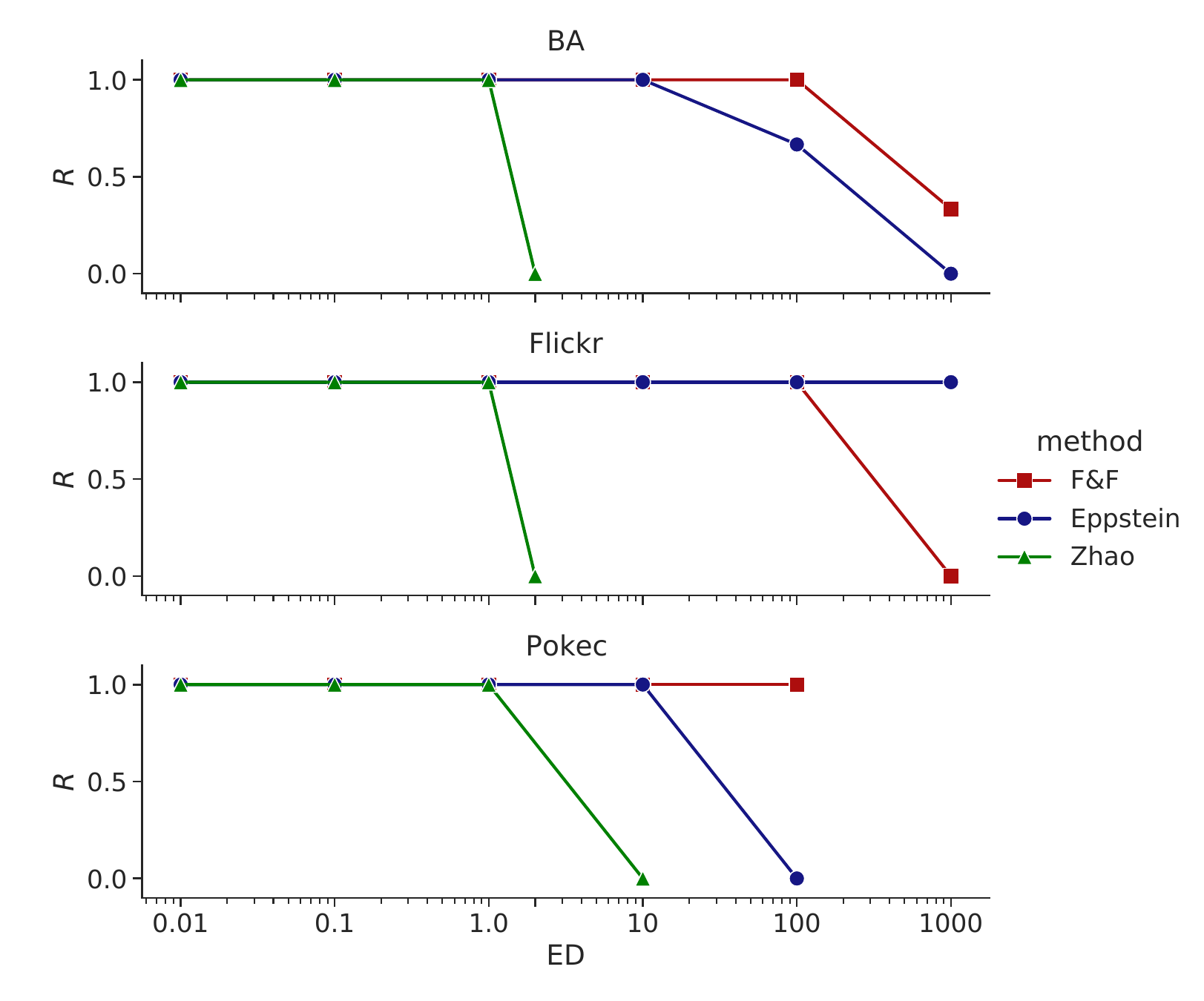}
 \caption{Robustness\label{fig:edgesflips}}
 \end{subfigure}
 \caption{Benchmark comparison between Scheme, Eppstein~\cite{isc}, and Zhao~\cite{COSN}. (a) Timing ($y$-axis) of the \texttt{Embed} and \texttt{Extract} functions for the three competing schemes, depending on the BA graph size $N$ ($x$-axis). Executions taking less than the timeout (fixed at 25 minutes) are kept and plotted. (b) Robustness of the three schemes ($y$-axis) facing attacks: the normalized number of successful extractions, facing increasing edge flips (measured by edit distance on the $x$-axis as ratio of edges) on three large graphs.}
 \label{fig:example}
\end{figure}

In Figure~\ref{fig:timings}, timings are measured in seconds as a function of the number of vertices $N$ in an increasing size BA graph (with an attachment parameter of $3$). We set a timeout of $25$ minutes for the experiments (dotted gray horizontal line), so that all scheme executions-- each running on a single core-- are discarded if they exceed this threshold. For small and medium-sized graphs ($N \leq 10^5$), the embedding function of Zhao \textit{et al.} is the most efficient method, while its extraction takes the longest. For medium and large graphs (more than $10^5$ vertices), the \scheme scheme outperforms the state-of-the-art. Without exceeding the timeout, \scheme can watermark graphs with $10$ times more vertices than the scheme of Eppstein \textit{et al.} and up to $100$ times more vertices than the scheme of Zhao \textit{et al.}\\
Extractions take place on a non-attacked $G_W$, which constitutes the best-case scenario for the methods proposed by Eppstein \textit{et al.} and Zhao \textit{et al.}: their extractions are longer when applied to attacked graphs, because the approximate subgraph matching is more complex (while the matrix representation in \scheme makes it time-invariant to attacks). The extractions of Zhao \textit{et al.} and Eppstein \textit{et al.} are linear in the number of vertices, while \scheme is almost constant after $10k$ vertices.

\subsection{Robustness to the Edge Flip Attack}
\label{ss:zhao}
Figure~\ref{fig:edgesflips} shows the results for the three large graphs. The $y$-axis is the success rate. The $x$-axis shows the number of edges flipped, measured by the ED. Each experience averaged three times (the standard deviation is zero, as there is no variation in the results across all the performed executions).

We observe that \scheme performs better than the scheme of Zhao \textit{et al.} for the three graphs. The scheme of Zhao \textit{et al.} fails before $10$\% edges flipped, while \scheme handles flips equivalent to $100$\% edges. \scheme also beats the Eppstein scheme, except on the Flickr graph where Eppstein survives $1000$\% of flips.

We conclude that \scheme is at least as robust-- except on one configuration--, if not better, against random edge flips than its two competitors. Overall, the high robustness of \scheme (resilience $\geq100$\% of flipped edges, see Table~\ref{expres} and Figure~\ref{fig:example}b) makes it practical even in scenarios with extreme attack strengths.

\section{Conclusion}
\label{Conclusion}

Large real graphs represent valuable assets that watermarking techniques can track for provenance. This paper bridges the gap between image processing and graph domains. While previous state-of-the-art schemes relied on standard graph manipulations,
we have shown that knowledge and techniques from the multimedia community can be ported to the graph domain, yielding significant complexity improvements. The process involves studying the transformations between a graph and its corresponding real-valued image-like representation, and vice versa. 
Our method \scheme shows strong performance on several metrics, including low false positives and negatives.

Future extensions of the framework may include other binarization methods, such as the optimized thresholding method proposed by Otsu~\cite{Otsu}. It may lead to even smaller watermarks, which is crucial for minimizing distortions in the watermarked graph. Another promising direction is to adapt recent image watermarking schemes to our framework.

\hspace{1cm}
\appendix

\subsection{Notations}\label{s:notations}
Table~\ref{notations} is a summary of our main notations. In particular, the differences between $\omega$, $\mathcal{W}$ and $W$ are indicated.

 \begin{table}[b!]
 \begin{center}
 \scalebox{1}{
 \begin{tabular}{ | c | c|}
  \hline
  $G$ & An arbitrary undirected and unweighted graph\\
   & with $N$ vertices $V$, and edge set $E$\\
  $G_W$ & The watermarked graph\\
  $G^*$ & Another arbitrary graph to test for the watermark presence\\
  $A$, $A_W$ & Adjacency matrices of graph $G$ and $G_W$ respectively\\
  $A'$ & Matrix obtained before the binarization of $A_W$\\
  $\theta$ &
  Similarity threshold for the watermark extraction success\\
  $\omega$ & Watermark key, of length $m$\\
  $\sigma$ & Standard deviation for generating the key\\
  $\chi$ & List of indices to insert the key\\
  $\mathcal{W}$ & Intermediary (not yet binarized) watermark\\
  $W$ & Final watermark\\
  \hline
 \end{tabular}
 }
 \end{center}
 \caption{Main notations in this paper.}
 \label{notations}
 \end{table}

\begin{figure}[h!]
\centerline{\includegraphics[width=0.5\textwidth]{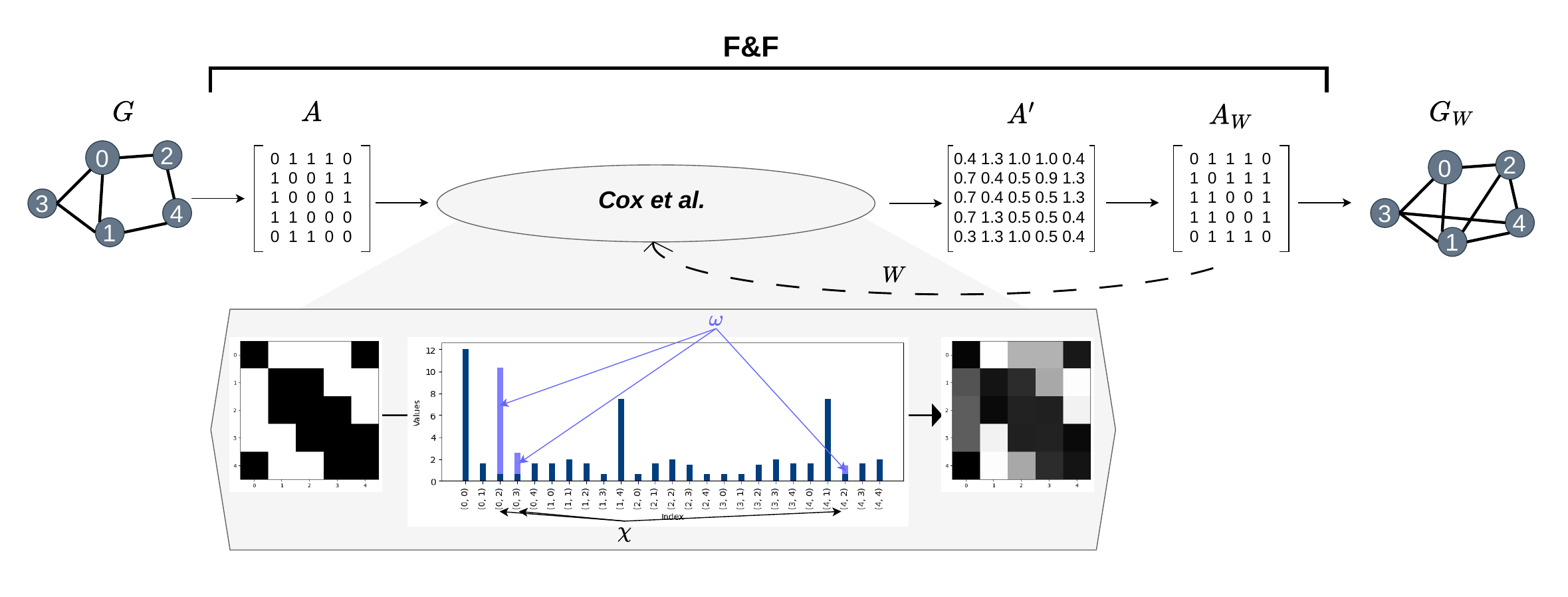}}
\caption{The \scheme framework for watermarking unweighted graphs using an image domain scheme (hereby Cox \textit{et al.}~\cite{TIP}).
} 
\label{fig:schema}
\end{figure}

\subsection{Watermarking non-graph objects}
\label{sec:related}

This section is an overview of methods for watermarking graph-related objects that are not graphs themselves.

\paragraph*{Watermarking various digital objects}
Initially introduced to protect multimedia objects~\cite{560423,TIP}, watermarking techniques are now being applied to a wide range of other areas, such as the model protection of deep neural networks \cite{le2020adversarial}, graph neural networks \cite{zhao2021watermarking} or large language models~\cite{kirchenbauer2023watermark} to just cite a few recent works. Watermarking approaches sometimes use graph abstractions to hide information in the object, as it is the case with software watermarking. (See e.g.~\cite{dey2019software} for a review.)
\par In software watermarking approaches, graph abstractions capture the interactions between variables of a source code to be watermarked. The purpose is thus to watermark the software; graph representations are simply leveraged to build watermarking schemes, and are not the target of the watermarks. Some approaches focus on databases, to watermark datasheets or RDBMS tuples, for example \cite{agrawal2002watermarking,cudre2011graph,kumar2020recent}; directly watermarking graph objects remains much more general, as targeting graph structures themselves \cite{COSN,isc}, and because no assumption is made on the presence or absence of tuples and rich data to in which embed information into.
\par In graph signal watermarking~\cite{icassp}, since signals are data defined on the vertices of graphs, it is natural to watermark them by using graph eigenvectors. However, this technique thus consists of watermarking signals defined over graphs, and it never watermarks graphs themselves.

\paragraph*{Watermarking graphs}
This paper is dedicated to watermark unweighted graphs.
Recently, two graph watermarking approaches have been introduced that use purely graph techniques~\cite{isc,COSN} to watermark graphs. Since they are direct competitors to \scheme, both schemes are described in detail in the main part of the paper (Section~\ref{sec:benchmark}), and then benchmarked.

\paragraph*{Digital image watermarking}
Efficient image watermarking has motivated decades of research. Images can be watermarked by taking into account features such as the perception of the human visual system (HVS)~\cite{HVS}, or by applying structural operations to them such as JPEG compression~\cite{JPEG}. Ingemar J. Cox \textit{et al.}~\cite{TIP} proposed the general method of Fourier transform and in-spectrum watermarking. This work is still used today for other types of transforms such as wavelets~\cite{wvl} or fractional Fourier transforms~\cite{FFT}.

\subsection{Theoretical Analysis of Uniqueness}\label{s:theory}
Let $G_{W_1}$ be the watermarked graph resulting from a pair $(G, \omega_1)$. $G$ is the graph to be watermarked and $W_1$ the resulting watermark from $\omega_1$. The uniqueness problem is to determine whether $G_{W_1}$ is unique. 

The only source of randomness in the scheme is the generation of the key; thus if $G$ is watermarked with another key $\omega_2$ resulting in $G_{W_2}$, then there is uniqueness if $G_{W_1} \neq G_{W_2}$. This property is not trivial because of the binarization which spreads keys over all the Fourier coefficients of the adjacency matrices. We then study the probability of uniqueness:
\begin{equation}\label{result}
    P(G_{W_1} \neq G_{W_2}) = 1 - P(G_{W_1} = G_{W_2})
\end{equation}

\paragraph*{Proof}
    Let $A_{W_1}$, respectively $A_{W_2}$, be the adjacency matrices of $G_{W_1}$, respectively $G_{W_2}$. By the hypothesis of the deterministic mapping of vertices into adjacency matrices:
$P(G_{W_1} = G_{W_2}) = P(A_{W_1} = A_{W_2}).$
Similarly, let $A'_1$, respectively $A'_2$ be the matrices corresponding to $A_{W_1}$, respectively $A_{W_2}$ \textbf{before} the binarization.

By definition, binarization is done using the average of the adjacency matrix $A$ of $G$:

\begin{center}
    \begin{flalign} \label{eq:P}
    &P(G_{W_1} = G_{W_2}) 
        = P\left(\bigcap_{i,j} (A_{W_1}[i,j] = A_{W_2}[i,j])\right)\nonumber\\
    & = P\left ( \bigcap _{i,j}\begin{matrix}
    | A'_{1,2}[i,j] | \leq av(A) \wedge A[i,j]=0\\ 
    \cup | A'_{1,2}[i,j]| > av(A) \wedge A[i,j]=1\\ 
    \cup |A'_{1,2}[i,j]| \leq av(A) \wedge A[i,j]=1\\ 
    \cup |A'_{1,2}[i,j]| > av(A) \wedge A[i,j]=0
     \end{matrix} \right ).
    \end{flalign}
\end{center}

Here the notation $A'_{1,2}$ means that the result must be true for both $A'_{1}$ and $A'_{2}$ respectively and $av(A)$ is the average of all elements of the matrix $A$. By the linearity of the inverse Fourier transform, for $n \in \{1,2\}: A'_n = A + FT^{-1}(\mathcal{W}_n).$

By construction, all the elements of each key $\omega$ (and its matrix version $\mathcal{W}$) are independent. The probability $P(G_{W_1} = G_{W_2})$ in Equation~\eqref{result} can be computed with the cumulative distribution of $FT^{-1}(\mathcal{W}_n)$ ($n \in \{1,2\}$).

First, note that: $FT^{-1}(\mathcal{W}_{n}) = Z \mathcal{W}_{n} Z,$
where $Z[k,l] = \frac{1}{N}e^{2\imath\pi kl/N}$. Thus, $FT^{-1}(\mathcal{W}_n)$ is a Gaussian matrix. With this, we can get an exact formula for the probability distribution function of this inverse Fourier transform flattened to one dimension~\cite{zdvi}.
In fact, let $\tilde{\mathcal{W}_{n}}$ be the flattened vector of $\mathcal{W}_{n}$, as $\forall k \in \{1,...,N^2\}, \tilde{\mathcal{W}_{n}}[k] = \mathcal{W}_{n}[\lceil k/N \rceil ,(k-1) \mod N+1],$
where $\lceil k/N \rceil$ being the upper integer part of $k/N$ and $(k-1) \mod N$ being the rest of the Euclidean division of $k-1$ by $N$. Using these notations, we easily show that:
$FT^{-1}(\tilde{\mathcal{W}_{n}}) = \tilde{Z} \mathcal{W}_{n},$
where $\tilde{Z} = Z \bigotimes Z$ is the $N^2 \times N^2$ Kronecker product of $Z$ with itself. Let $(z_{k,l})$ be the coefficients of $Z$, then for $(k',l') \in \{1,...,N^2\}$: $(\tilde{Z})_{k', l'} = z_{\lceil k' /N\rceil , \lceil l' /N\rceil} z_{(k'-1) \mod N+1, (l'-1) \mod N+1}.$ Since $\tilde{\mathcal{W}_{n}}$ is a Gaussian vector, $FT^{-1}(\tilde{\mathcal{W}_{n}})$ is also a Gaussian vector. Its average and standard deviation are computed using the matrix $\tilde{Z}$. In conclusion, the watermark's uniqueness is calculable.

\paragraph*{Application on large graphs.} The connection between graph densities and collisions is given by the theoretical analysis in Equation~\eqref{eq:P}. Indeed, as $av(A) = \frac{|E|}{N^2}$, the threshold for binarization is equal to the density divided by $N$. For the $N=1M$ graphs (density less than $10$) in Figure~\ref{fig:uniqueness} and the given keys:
$av(A) \simeq 0,$
$av(|FT^{-1}(W)|) \simeq 0,$ and
$av(|A+FT^{-1}(W)[i,j]|) \simeq av(A).$

Then, it is clear that the denser a graph, the more $A[i,j]$ coefficients are equal to $1$, and decreases the probability of collision (as seen in Equation~\eqref{eq:P}).

\paragraph{Successful Embeddings: a Corollary of Uniqueness}
In this section, we show that studying a successful embedding is a corollary of the analytical form of the uniqueness. The embedding of a key $\omega$ in a graph $G$ is successful if the watermarked graph $G_W$ is different from the original graph $G$. It can happen that the binarization discards a key built with insufficiently large parameters, so that no watermark is inserted into the graph. The study of successful embedding is therefore the estimation of $P(G_{W} = G)$.

\paragraph*{Proof}
Since $\texttt{Keygen}(0,0) =  []$ (an empty key) and  $\texttt{Embed}(G, []) = G$, studying $P(G_{W} = G)$ is a special case of Equation~\eqref{result} where $G_{W_1} = G_{W}$ (the graph watermarked with the key $\omega$) and $G_{W_2} = G$ (the graph watermarked with the empty key $[]$).

Note that in practice, the owner of the graph knows when the embedding has succeeded or failed: if the watermarked graph is equal to the original graph after an embedding attempt (which is linear to check in $O(|E|)$), it is sufficient for the owner to retry an embedding operation with larger key parameters until the two graphs are truly different.

\subsection{Scalability reduction}\label{a:reduction}
In Section~\ref{ss:scalability}, we explain that a scability reduction is applied to \scheme as it is the case in ~\cite{isc, COSN}. This section provides the details and explanation of such an operation. 

When considering graphs, the most significant information is contained in the vertices of the highest degree vertices; thus, analogous to low frequencies, in \scheme, the watermark is inserted into the lowest amplitude coefficients in the spectrum among the higher-degree vertices.
While the size of images on which the Cox \textit{et al.} scheme is typically applied is in the order of tens of thousands of pixels,
we also choose to take the top $N_0 = 10^4$ higher-degree vertices of $G$ to apply \scheme.

In this way, \texttt{Embed} and \texttt{Extract} are applied only on $N_0$ vertices and not on the $N$ vertices of the original graph. Their worst-case time complexities, computed in Section~\ref{ss:complexity}, are now $\mathcal{O}\left( N_0^2 \log N_0\right)$. However, to determine the subgraph on which to apply \scheme, it is necessary to select the $N_0$ vertices of highest degrees (which is calculated in $\mathcal{O}\left( N \log N_0\right)$). This cost is added naturally to the \texttt{Embed} function. 

The worst-time complexities of \scheme, with the dimensionality reduction, are summarized in Table~\ref{complexities}.

\begin{table}[]
\begin{center}
\begin{tabular}{ccc}
\scheme & raw complexities & adapted complexities\\
 \hline
\texttt{Keygen} & $\mathcal{O}(m)$ & $\mathcal{O}(m)$\\
\texttt{Embed} & $\mathcal{O}\left( N^2\log N\right)$ & $\mathcal{O}\left( (N + N_0^2) \log N_0\right)$ \\
\texttt{Extract} & $\mathcal{O}\left( N^2 \log N\right)$ & $\mathcal{O}\left( N_0^2 \log N_0\right)$
\end{tabular}
\end{center}
 \caption{Worst-time complexities of \scheme to watermark a graph $G$ with $N$ vertices with a key of length $m$ with and without the dimensionality reduction to $N_0$ vertices.
 }
\label{complexities}
\end{table}

\paragraph{Guideline's complexities}
In Section~\ref{ss:distorsion}, we give practical guidelines for low distortion. We now studied this operation in detail, and its effect on the scalability of \scheme.

As mentioned before, we set the key length $m$ of \scheme as the key length used in~\cite{isc}. In fact, $m$ can be set in worst-case complexity $\mathcal{O}(N_0)$. Once the key length $m$ is set, it is always possible to find a value $\sigma$ to watermark the graph while achieving the distortion goal. Experiments have shown that for any graph $G$ and for any fixed key length $m \in [\![1, N_0*(N_0-1)/2]\!] $, there exists a value $\sigma_{max}$ to watermark the graph with a strictly positive edit distance. $\sigma_{max}$ can be estimated with at most $\mathcal{O}(\log N_0)$ operations.  An automated way to set $\sigma$ resulting in a small ED guarantee, is a dichotomous search in $[\![1, \sigma_{max}]\!]$. The worst-case time complexity of such setting is testing $\mathcal{O}(\sigma_{max} \log \sigma_{max})$ values for $\sigma$ when each test is of complexity $\mathcal{O}\left( N_0^2 \log N_0\right)$. To summarize, the worst-case time complexity for setting $\sigma$ knowing $m$ is 
$\mathcal{O}\left( N_0^2 \log N_0\right)$. This is the same as the worst-case complexity of the embedding function, so the proposed guidelines do not affect the scalability of \scheme.

\subsection{Additional experiments}\label{a:add-exp}
\paragraph{(Goal 5) Testing undetectability w. Graph Neural Networks}
In addition to the properties stated in the paper and which follow the properties stated in the related papers, we add a Goal 5. We want to measure whether a graph neural network is able to distinguish a watermarked graph from an original graph. Indeed, if such a network existes, it could be maliciously used by an adversary.

We present a novel attack strategy. So far, the approach studied has been to modify the graph in an attempt to remove the watermark. While the two related approaches assume that their scheme is undetectable by construction, we decide to go a step further and propose a new attack that focuses on identifying the watermarked graphs instead. If successful, this implies that the watermark significantly degrades the graph despite a small ED. To achieve this, we use a conventional Graph Neural Network (GNN) that solves a graph classification task. To train a predictor for identifying watermarked graphs, we first need a labeled dataset. 

We generated $2,000$ independent Barabási-Albert graphs of $1,000$ vertices with an attachment parameter of $3$. Subsequently, we watermarked half of these graphs by generating $1,000$ independent keys with parameters $m = 200$ and $\sigma = 100$ (chosen to achieve an ED between $1$ and $100$ edges of the graph, i.e. between $0.034$ and $3.4$\% edges). 
Our dataset includes the $1,000$ non-watermarked graphs (of label 0) and the $1,000$ watermarked graphs (of label 1). Notably, we underline that the seed graphs for generating the watermarked graphs were \textbf{excluded} from the dataset, to avoid bias.

We adapted the graph classification tutorial from the Python library called PyTorch Geometric~\cite{GNNtuto} with our dataset. The task of the GNN is to distinguish graphs into two classes: those who have been watermarked and those who were not. The degrees of the vertices have been added as their features in this dataset. In other words, the degree sequence of each graph will serve as a basis for attempts in detecting a deviation from the benign sequences of non watermarked graphs.
The training is set to $80$\% of the dataset, leaving $20$\% as a test set.

Experiments show that with its default parameters or some classic variants of these (number of hidden channels, learning rate...), the GNN does not succeed in the task of discriminating between watermarked graphs and unwatermarked graphs. Indeed the accuracy on train and test datasets oscillates between 0.4775 and 0.5225 depending on hyperparameters.

We conclude that this attack is ineffective against \scheme. This implies that the watermarks inserted in \scheme are small enough (ED between $0.034$ and $3.4$\% edges) to be undetectable by such a GNN classification attack.

\paragraph{Testing Laplacian eigenvalue's stability}
In our examples, we focus on Edit Distance (ED) due to its generality and widespread use in the state-of-the-art. We also consider degree centrality, as shown in Figure~\ref{fig:spearman}. The next section discusses the impact of \scheme on the Laplacian eigenvalues of a graph, a metric that is sometimes used in graph studies. 

There is no theoretical guarantee of the invariance of the Laplacian eigenvalues after applying \scheme. However, we performed additional experiments with Barabasi-Albert graphs of different sizes ($1k$ or $10k$ nodes) and densities (attachment parameter: $3$, $5$, or $7$). Experimentally, the maximum variation of an eigenvalue is at most $\pm-4\%$ of the initial value. However, there is no conservation of the smallest non-zero eigenvalue, which can vary by up to $2\%$ of its value (e.g., from $\sim 1.268$ to $\sim 1.297$ for a BA graph with $10k$ vertices and $70k$ edges). Depending on the intended use of the graph, this variation of up to $4\%$ over the entire spectrum (calculated as an infinite norm) may be acceptable.

In conclusion, we suggest that future readers test their own metrics of interest before sharing graphs watermarked with \scheme. Even if there is no theoretical guarantee on the preserving of this metric, the variation induced for the watermark is sometimes negligible for the desired use.

\bibliographystyle{plain}
\bibliography{references}

\end{document}